\documentclass[a4paper,12pt]{article}
\usepackage{graphicx}
\usepackage{epsfig}
\usepackage[american]{babel}
\usepackage{amssymb}
\usepackage{calc}

\begin{document}

\title{Phenomena of complex analytic dynamics
in the non-autonomous, nonlinear ring system}
\author{O.B.~Isaeva$^{1*}$, S.P.~Kuznetsov$^1$, M.A.~Obichev$^2$}
\date{}
\maketitle\begin{center} \emph{
$^1$Kotel'nikov Institute of Radio-Engineering and Electronics of RAS, Saratov Branch \\
Zelenaya 38, Saratov,
410019, Russian Federation}\end{center}

\maketitle\begin{center} \emph{$^2$Saratov State University
\\ Astrahanskaya 83 Saratov,
410026, Russian Federation}\end{center}

\maketitle\begin{center} \emph{$^*$}IsaevaO@rambler.ru\end{center}

\begin{abstract}
The model system manifesting phenomena peculiar to complex analytic 
maps is offered. The system is a non-autonomous ring cavity with 
nonlinear elements and filters,

\end{abstract}

\maketitle

When one studies simple quadratic map of a complex variable and with 
complex parameter
\begin{equation} \label{pe1}
z_{n+1} = c + z_{n}^{2}.
\end{equation}
the interesting phenomena can be observed. The complex phase plane of 
a variable $z$, appears subdivided to two domains. The start from one 
area gives trajectory $z_n$, escaping to infinity. The start from another 
area gives trajectory, remaining restricted: the trajectory wanders 
not far away from an origin. This second domain is named 
a Julia set. In a trivial case with $c=0$ it looks as a unit disk. 
With nonzero parameter values the Julia set is, fractal with non-smouth 
border, and has self-similar structure (see Fig. 1).

\begin{figure}
\centerline{\epsfig{file=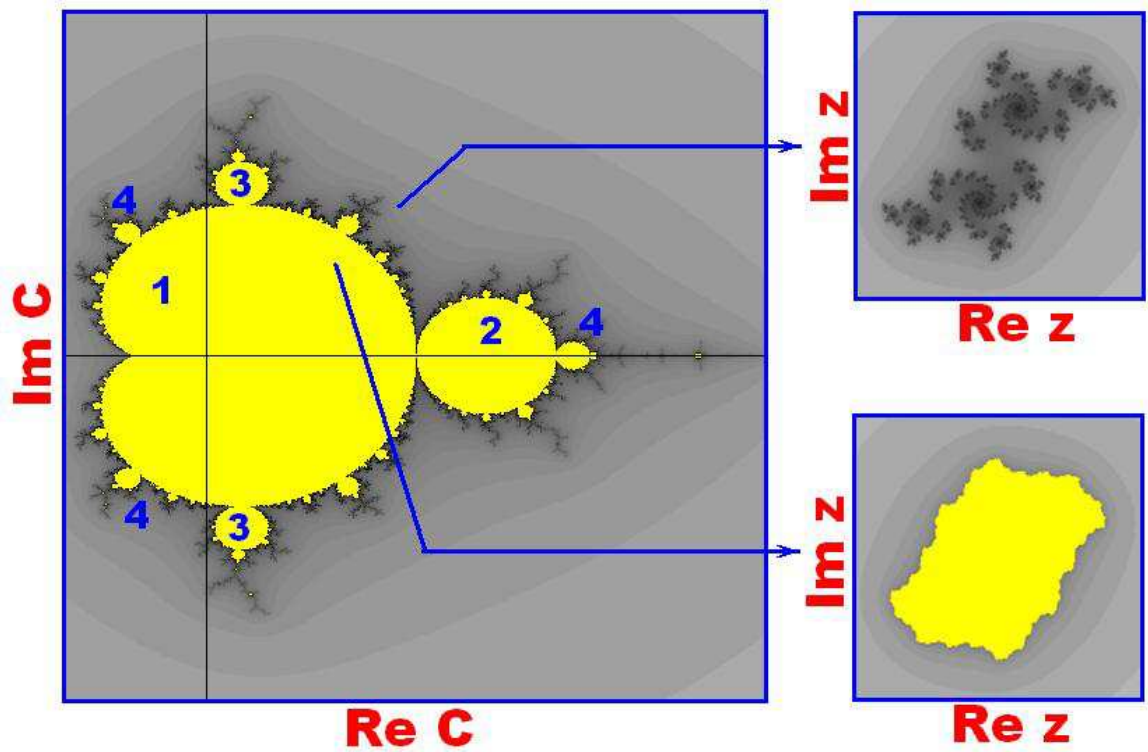,bb = 20 20 592 395,width=0.9\textwidth}}
\caption{Mandelbrot set~(left panel) and Julia sets (right panels) for the 
complex quadratic map~(1) at parameter and phase plane charts. Yellow areas 
correspond to the periodic dynamics of the complex variable 
$z$ in a bounded domain (the periods are marked by figures). Black color 
designates bounded chaotic dynamics. 
Gray colors correspond to the escape of the trajectories to infinity
by different steps of time (more dark color corresponds to more long escape 
time)}
\end{figure}

 The restricted in a phase space dynamics is possible not with any parameter 
$c$ values. Moreover, with different parameter values it has different 
properties -- it can be periodic or chaotic. The useful method for examination 
of a system is the drowing of a chart of a complex parameter plane, for example 
as a chart of dynamical regimes. At Fig.~1 such chart is represented. On the plane 
$c$ the object known as a Mandelbrot set is arises. It represents a fractal 
domain, parameter values from which corresponds to restricted dynamics. 
Mandelbrot set looks 
as a cardioid with a set of the attached round lobes (corresponded to periodic 
dynamics). This "cartus" is enclosed by fractal "mane", corresponded to chaotic 
dynamics.

The huge amount of the mathematical literature is devoted to Mandelbrot and 
Julia sets [1-3]. The important task for physicist is the development of 
physical applications of the theory of complex analytic dynamics and 
building up of the physical devices and systems in which the complex 
dynamics is implemented. Just several examples of such systems are exists [4-9]. 
In the present work a new example of a system in which Mandelbrot set can 
be implemented is offered. It is a non-autonomous ring system with nonlinear 
elements and filters. This model can be constructed for example on the basis 
of an optical system of the Ikeda cavity type [10] or as a ring system with the 
ferromagnetic structures as nonlinear elements [11].

Let us consider the system represented at Fig.~2. The signal can travel 
between knots 1-2-3-4. In the optical resonator these are the mirrors, 
turned around from each other so that a ray of light transits on a ring 
path. An external pumping C and D exciting the signal in the system are 
brought at knots 1 and 3. In optical cavity it can be semi-glassy mirrors. 
Signals C and D have complex slow amplitude $C$ and $D$ and frequency $\omega$.

\begin{figure}
\centerline{\epsfig{file=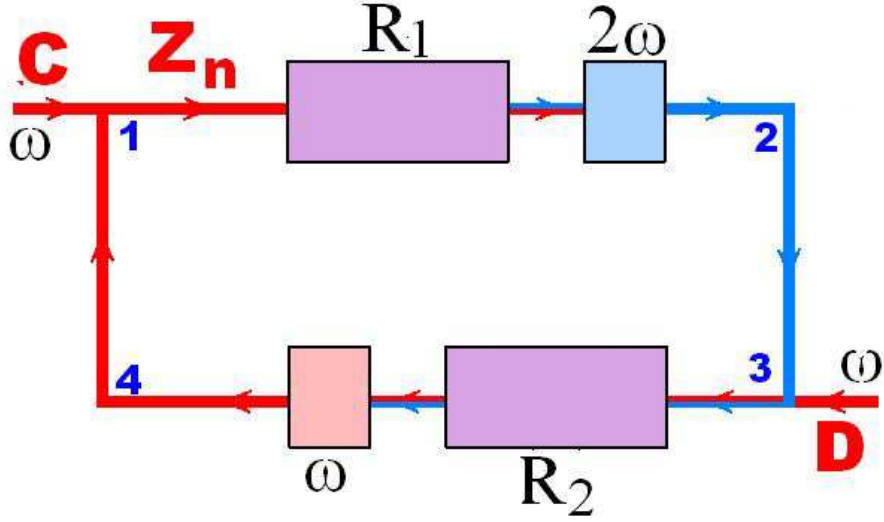,bb = 20 20 592 359,width=0.7\textwidth}}
\caption{Diagram of the ring device with two nonlinear elements (violet 
rectangles of length $R_1$ and $R_2$), two filters, absorbing the signal of frequency 
$2\omega$ (blue box) and the signal of frequency $\omega$ (pink box).
C and D are the external pumping signals of frequency $\omega$. $Z_n$ is a 
complex slow amplitude of the signal (of frequrncy $\omega$) near after the 
knot, marked as 1}
\end{figure}

Let us assume that near the knot 1 signal in the ring have slow complex 
amplitude $Z_n$ and frequency $\omega$. By travelimg (the direction is 
indicated by arrow) a signal  spreads through the segment of nonlinear 
medium of length $R_1$, in which the signal transforms. For example the 
component on doubled frequency. can be excited. Interaction of two these 
components (in the case of its synchonism and by approximation of 
statiobarity of waves) can be described by differential equations
\begin{equation} \label{pe2}
i\frac{\partial a}{\partial x}=\alpha a^*b, \qquad i\frac{\partial b}{\partial x}=\beta a^2
\end{equation}
where $a(x)$ is a complex slow amplitude of the component on frequency 
$\omega$, $b(x)$ is a complex slow amplitude of a component on frequency 
$2\omega$, $x$ is a coordinate along a segment of nonlinear medium.
If on an input to a medium the amplitude of component with frequency 
$\omega$ is $a(0)\sim Z_n$, then on an exit in some approximation the 
amplitude of the component of frequency $2\omega$ is $b(R_1)\sim Z_n^2$.

Further a signal runs through a filter which passes only component on 
frequency $2\omega$. After that the signal is reflected from a knot 2 
and a knot 3, where external pumping D of frequency $\omega$ is added. 
On a segment between the third and fourth knot the nonlinear medium of 
length $R_2$ is located. Transiting through nonlinear medium 
signal (with a component of frequency $2\omega$ come from the filter 
and a component of frequency $\omega$ added outside) transforms according 
to the equations (2). On an exit from a nonlinear medium the component of 
a signal on frequency $\omega$ in some approximation have intensity 
$a(R_2)~D ^*Z_n^2$.

Further the signal transits through the filter which passes component 
on frequency $\omega$. Finally, on a knot 1 there is an additional 
external pumping C on the same frequency. Complex amplitudes of external 
signal and signal in the ring accumulates $a(R_2)+C$. Thus, the signal 
makes the full circle. It have the same frequency $\omega$ as in the 
beginning of the path and its complex amplitude can be asymptotically 
wrote as $Z_{n+1}=F(Z_n)=C+D^* Z_n^2$. Last expression represents complex 
map.

Numerical experiment of evolution of a signal in this system is carried out. 
The equations (2) are integrated (by Runge-Kutta method). As a matter of fact 
the numerical simulation is divided into some stage:  1) an integration of a 
system (2) with parameters $\alpha_1$, $\beta_1$ and with initial conditions 
$a(0)=Z_n$, $b(0)=0$ gives us $a(R_1)$ and $b(R_1)$; 2) an integration of a 
system (2) with parameters $\alpha_2$, $\beta_2$ and with initial conditions 
$a(0)=D$, $b(0)=b(R_1)$ gives us $a(R_2)$ and $b(R_2)$; 3) accumulation 
$a(R_2)+C$ gives us a new step value of a complex discrete variable $Z_{n+1}$.
At Fig.~3 two charts of a plane of parameter $C$ for an investigated 
system at different values of parameters, characterising the nonlinear media 
are represented. 

Obviously the Poicar{\'e} map $Z_{n+1}=F(Z_n)$ is not complex analytical (it is not 
satisfies Cauchy--Riemann equations) because system (2) contains non-analytical 
term, proportional to $a^*$. Phenomena of complex analytic dynamics can 
realise in described model only as a approximation.  
One can see at Fig.~3  "almost ideal" Mandelbrot set on top panel 
which is distorted on bottom panel by the growth of parameters $\alpha_1$ 
and $\beta_2$ (leading to amplification of "non-analyticity").

\begin{figure}
\centerline{\epsfig{file=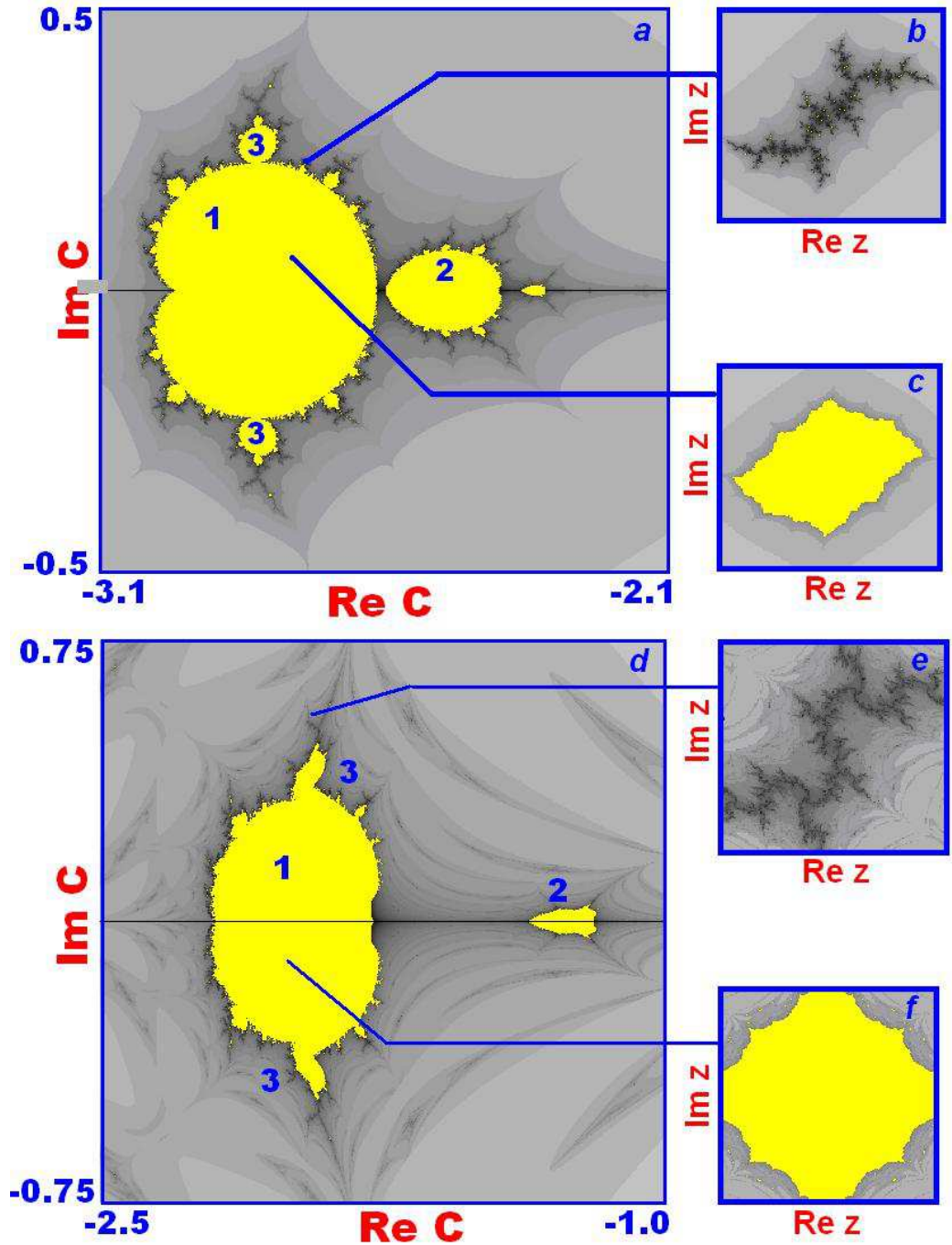,bb = 20 20 583 772,width=0.8\textwidth}}
\caption{Mandelbrot-like set~($a$,~$d$), which arises for the ring cavity represented 
at Fig.~2, on the chart of the complex plane of a parameter $C$, responsible 
to the complex slow amplitude of pumping signal C, and Julia-like 
set~($b$-$c$,~$e$-$f$), which arises at he complex plane of the amplitude of 
the signal $Z$ in the ring in the Poincar{\'e} section near knot 1.
Values of parameters are: complex slow amplitude of the pumping signal D 
$D=3$; length of nonlinear elements $R_1=R_2=1.0$; parameters, characterising
media in nonlinear elements $\alpha_1=0.01$, $\beta_1=1.0$, $\alpha_2=1.0$,  
$\beta_2=0.01$ ($a$-$c$) and $\alpha_1=0.05$, $\beta_1=1.0$, $\alpha_2=1.0$, 
$\beta_2=0.1$ ($d$-$f$)}
\label{pf3}
\end{figure}

One tool for an analysis of the degree of violation of the complex
analyticity, suggested in~\cite{pl12} , is the computation of the spectrum of Lyapunov
exponents. In particular, for a two-dimensional map
$X_{n+1}=U(X_n,Y_n)$, $Y_{n+1}=V(X_n,Y_n)$ they may be determined
via the eigenvalues of the matrix
\begin{equation}\label{pe481}
\mathbf{b}=\mathbf{a}(X_0,Y_0)\mathbf{a}^+(X_0,Y_0)
\mathbf{a}(X_1,Y_1)\mathbf{a}^+(X_1,Y_1)...
\mathbf{a}(X_{M-1},Y_{M-1})\mathbf{a}^+(X_{M-1},Y_{M-1}),
\end{equation}
where
\begin{equation}\label{pe48}
\mathbf{a}=\left(
\begin{array}{cc}
  \partial U(X,Y)/\partial X & \partial U(X,Y)/\partial Y \\
  \partial V(X,Y)/\partial X & \partial V(X,Y)/\partial Y
\end{array}
\right),
\end{equation}
In the case of a two-dimensional real map equivalent to an
analytic map of one complex variable, two Lyapunov exponents must
be equal. It may be shown from the Cauchy-Riemann conditions 
\begin{equation}\label{pe3}
\begin{array}{cc}
  \partial U(X,Y)/\partial X = \partial V(X,Y)/\partial Y, \\
  \partial U(X,Y)/\partial Y = - \partial V(X,Y)/\partial X,  
\end{array}
\end{equation}
that two eigenvalues coincide at any values of parameters and
variables. The same is true for the Lyapunov exponents expressed
as $\Lambda_{1,2} \sim \log\lambda_{1,2}$.

The studying ring system possesses four Lyapunov
exponents. To compute the Lyapunov exponents we used the Benettin
algorithm~\cite{pl13}. The procedure consists in simultaneous
numerical solution of the equations~(2) and a collection
of four exemplars of the linearized equations for small
perturbations:
\begin{equation} \label{pe4}
i\frac{\partial \widetilde{a}}{\partial x}=\alpha \widetilde{a^*}b + \alpha a^*\widetilde{b}, 
\qquad 
i\frac{\partial \widetilde{b}}{\partial x}=2\beta a \widetilde{a}
\end{equation}
with parameters $\alpha_1$, $\beta_1$ during $x\in(0,R_1)$. and 
with parameters $\alpha_2$, $\beta_2$ during $x\in(0,R_2)$. Signal 
transformation in the filters are taken into account as the shift of the 
variables and the perturbation vectors. Pumping signal C gives only impact to 
the variable.

At each Poincar{\'e} section after passing knot 1 in the ring 
we perform Gram-Schmidt
orthogonalization and normalization for a set of four vectors
$\widetilde{\mathbf{x}}^j=\{\widetilde{\mathrm{Re} a}^j,
\widetilde{\mathrm{Im} a}^j, \widetilde{\mathrm{Re} b}^j,
\widetilde{\mathrm{Im} b}^j\}$,~$j=1,...,4$. The Lyapunov
exponents are estimated as mean rates of growth or decrease of
logarithms of the norms of these four vectors:
\begin{equation}\label{pe50}
\Lambda_j=\frac{1}{M}\sum_{i=1}^{M}\ln\|\widetilde{\mathbf{x}}_i^j\|,
\qquad j=1,...,4,
\end{equation}
where the norms are evaluated after the orthogonalization but
before the normalization.

Computations show that, depending on the regime, two larger
exponents may be negative (periodic attractive orbits), positive
(chaotic motions) and zero (a border of chaos and quasiperiodic
regimes). The second Lyapunov exponent is more or less close to 
the first one.
The other two exponents are always negative in the whole
domain of existence of bounded dynamical (i.e. on the
Mandelbrot set) and tend $-\infty$. 
In the left column of Fig.~4 we present
charts of the largest Lyapunov exponent on the
plane~$C$ for the same parameters values as at Fig.~3. 
Gray tones from light to dark correspond to variation of
the Lyapunov exponent from~$0$~to~$-\infty$. Observe that at central 
parts of the Mandelbrot set leaves the largest Lyapunov exponent 
becomes large negative, which
corresponds to periodic motions of high stability. At edges of the
leaves, a thin strip of appearance of positive Lyapunov exponent
can happen (chaos). The picture is similar to that for the
quadratic complex analytic map; see~e.g.~Ref.~\cite{pl14}. At panel ($c$), 
distortion of the configuration develops. For example, one can see thick 
stripe of chaos. The leaves lose their round form and separate each other. 
In the right column of Fig.~4, we depict respective charts for the 
difference of the two larger 
Lyapunov exponents. The regions of large difference of the exponent 
are visualized by dark gray and black color. It reveals the
essential deviation from complex analytic dynamics.

\begin{figure}
\centerline{\epsfig{file=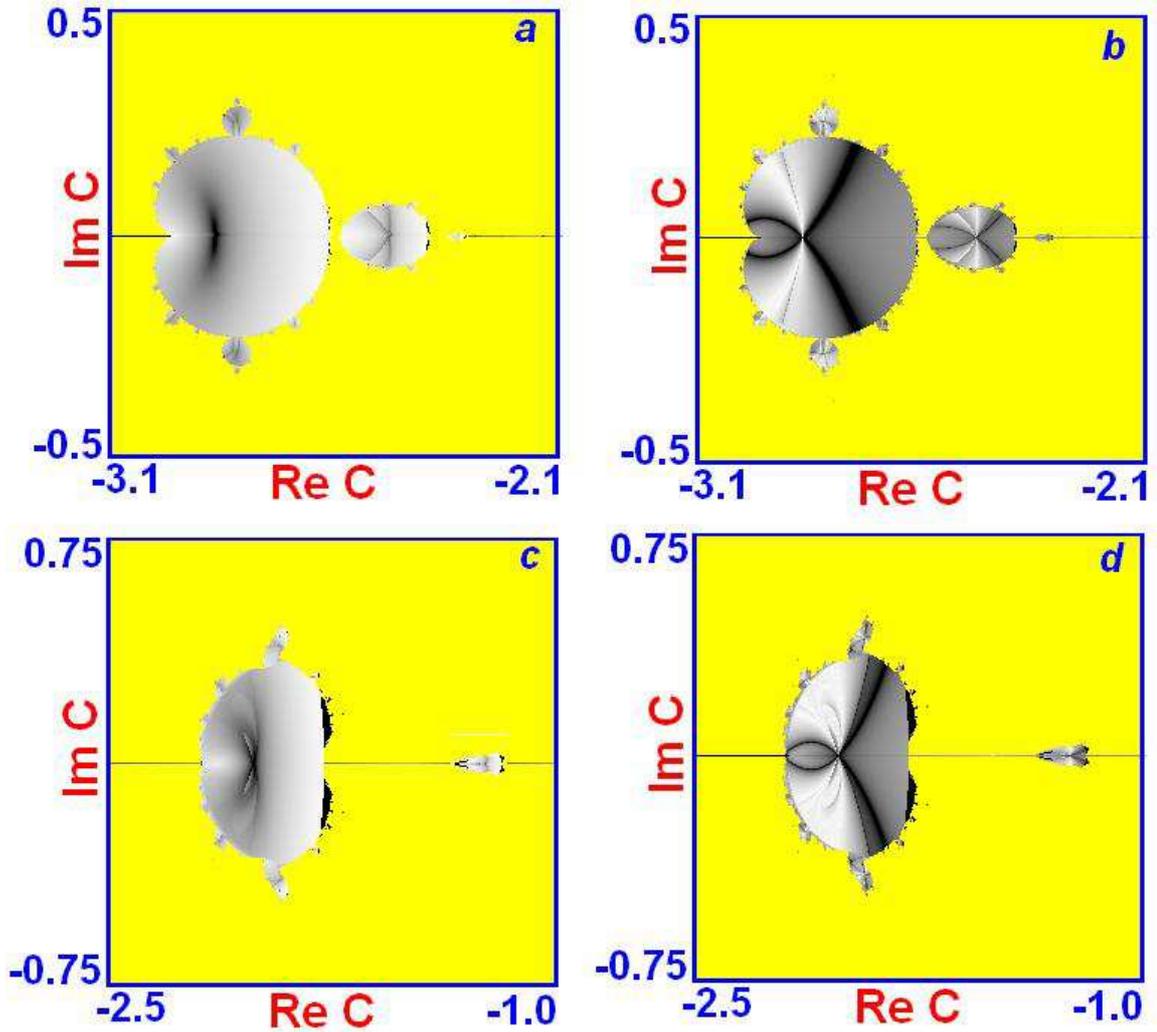,bb = 20 20 592 533,width=0.9\textwidth}}
\caption{Charts of the largest Lyapunov exponent ($a$, $c$) and for the
difference of the two larger exponents ($b$, $d$) for the ring system 
at the values of parameters, responsible for nonlinear mediae properties: 
$\alpha_1=0.01$, $\beta_1=1.0$, $\alpha_2=1.0$, $\beta_2=0.01$ ($a$, $b$) 
and $\alpha_1=0.05$, $\beta_1=1.0$, $\alpha_2=1.0$, $\beta_2=0.1$ ($c$, $d$). 
Values of other parameters are: $D=3$; $R_1=R_2=1.0$.
Uniform color means area of unstable dynamics (divergence to infinity). 
Black color on the diagrams corresponds to chaotic dynamics in a bounded 
domain. More dark tone of gray color in the left column corresponds to 
smaller value of the senior Lyapunov exponent and more strong stability 
of the periodic regime. More dark tone of gray color in the right column means 
larger difference brtween Lyapuniv exponents and therefor more considerable 
distortion of the complex analyticity.} \label{pf4}
\end{figure}

\newpage

\section*{Acknowledgement}
The work is performed under support from Grant of the
President of Russian Federation (MK-905.2010.2), RFBR (grant No. 11-02-00057)
and Federal special programm "Scientific and pedagogical personnel of innovative 
Russia" of 2009-2013 (project No. 2010-1.2.2-123-019-002).

newpage
\newpage

\end{document}